\documentstyle[prb,aps]{revtex}
\begin{document}
\draft

\title{Efficient index handling of multidimensional 
       periodic boundary conditions}

\author{Jos\'e M.\ Soler}

\address{Departamento de F\'{\i}sica de la Materia Condensada,
Universidad Aut\'onoma de Madrid, E-28049 Madrid, Spain.}

\maketitle

\begin {abstract}
   An efficient method is described to handle mesh indexes in  
multidimensional problems like numerical integration of partial 
differential equations, lattice model simulations, and determination 
of atomic neighbor lists.
   By creating an extended mesh, beyond the periodic unit cell, the
stride in memory between equivalent pairs of mesh points is independent
of their position within the cell.
   This allows to contract the mesh indexes of all dimensions into a
single index, avoiding modulo and other implicit index operations.
\end {abstract}


%
%
%
%

   Periodic boundary conditions are essential in all sorts of problems
in solid state physics, condensed matter simulations, and many other fields.
   The solution of partial differential equations by real-space 
discretization~\cite{Beck,NR}, the calculation of the interaction energy 
in lattice models~\cite{Binder}, and the determination of neighbor lists in 
molecular dynamics simulations~\cite{Allen-Tildesley},
are only a few of these problems.
   For illustration, let us consider the calculation of the laplacian 
of a function $f({\bf r})$ using finite differences.
   In three dimensions, one generally discretizes space in
all three periodic directions, using an index for each direction.
   For simplicity, let us consider an orthorhombic unit cell, with 
mesh steps $\Delta x, \Delta y, \Delta z$.
   Then the simplest formula for the laplacian is
\begin{eqnarray}
  \nabla^2 f_{i_x,i_y,i_z} &=& 
      (f_{i_x+1,i_y,i_z} - 2 f_{i_x,i_y,i_z} + f_{i_x-1,i_y,i_z}) / \Delta x^2
     \nonumber \\
  &+& (f_{i_x,i_y+1,i_z} - 2 f_{i_x,i_y,i_z} + f_{i_x,i_y-1,i_z}) / \Delta y^2
     \nonumber \\
  &+& (f_{i_x,i_y,i_z+1} - 2 f_{i_x,i_y,i_z} + f_{i_x,i_y,i_z-1}) / \Delta z^2
     \nonumber
\end{eqnarray}
   A direct translation of this expression into Fortran90 code might read
\begin{verbatim}
   Lf(ix,iy,iz) = - f(ix,iy,iz) * (2/dx2+2/dy2+2/dz2)                 &
      + ( f(modulo(ix+1,nx),iy,iz) + f(modulo(ix-1,nx),iy,iz) ) / dx2 &
      + ( f(ix,modulo(iy+1,ny),iz) + f(ix,modulo(iy-1,ny),iz) ) / dy2 &
      + ( f(ix,iy,modulo(iz+1,nz)) + f(ix,iy,modulo(iz-1,nz)) ) / dz2
\end{verbatim}
where the indexes $i_\alpha$ ($\alpha=\{x,y,z\}$) of the arrays 
{\tt f} and {\tt Lf} run from 0 to $n_\alpha-1$, as in C.
   There are two problems with this construction. 
   First, the {\tt modulo} operations are required to bring the indexes back 
to the allowed range $[0,n_\alpha-1]$.
   And second, the use of three indexes to refer to a mesh point implies
implicit arithmetic operations, generated by the compiler, to translate
them into a single index giving its position in memory.

   A straightforward solution to these inefficiencies would be to create
a neighbor-point list \verb'j_neighb(i,neighb)', of the size of the number
of mesh points times the number of neighbor points.
   However, although the latter are only six in our simple example, they 
may frequently be as many as several hundred, what generally makes this
approach unfeasible.
   A partial solution, addressing only the first problem, is to create six
(or more for longer ranges) one-dimensional tables
$j_\alpha^{\pm 1}(i_\alpha) = \mbox{mod}(i_\alpha \pm 1,n_\alpha)$
to avoid the modulo computations~\cite{Binder}.
   Here, I describe a multidimensional generalization of this method, 
which solves both problems at the expense of a very reasonable amount 
of extra storage.

   The method is based on an {\em extended mesh}, which extends
beyond the periodic unit cell, by as much as required to cover all the space
that can be reached from the unit cell by the range of the interactions
or the finite-difference operator.
   The extended mesh range is
$i_\alpha^{min}=-\Delta n_\alpha$ and 
$i_\alpha^{max}=n_\alpha - 1 + \Delta n_\alpha$, where 
$\Delta n_\alpha = 1$ in our particular example, in which the laplacian 
formula extends just to first-neighbor mesh points.
   In principle, in cases with a small unit cell and a long range, the mesh 
extension may be larger than the unit cell itself, extending over several
neighbor cells.
   However, in the more relevant case of a large system, we may expect
the extension region to be small compared to the unit cell.
   We then consider two combined indexes, one associated to the normal
unit-cell mesh, and another one associated to the extended mesh
$$
  i = i_x + n_x i_y + n_x n_y i_z, 
$$
$$
  i_{ext} = (i_x-i_x^{min}) + n_x^{ext} (i_y-i_y^{min})
           + n_x^{ext} n_y^{ext} (i_z-i_z^{min}), 
$$
where 
$n_\alpha^{ext}=i_\alpha^{max}-i_\alpha^{min}+1=n_\alpha+2\Delta n_\alpha$.
   The key observation is that, if $i_{ext}$ is a mesh point {\em within}
the unit cell ($0 \le i_\alpha \le n_\alpha-1$), 
and if $j_{ext}$ is a neighbor mesh point (within its interaction range, 
i.e. $| j_\alpha - i_\alpha | \le \Delta n_\alpha$), then the 
arithmetic difference $j_{ext}-i_{ext}$ depends only on the relative
positions of $i_{ext}$ and $j_{ext}$ (i.e. on $j_\alpha-i_\alpha$), 
but not on the position of $i_{ext}$ within the unit cell.
   We can then create a list of neighbor strides $\Delta ij_{ext}$, 
and two arrays to translate back and forth between $i$ and $i_{ext}$.
   One of the arrays maps the unit cell points to the central region of
the extended mesh, while the other one folds back the extended mesh points 
to their periodically equivalent points within the unit cell.
   Then, to access the neighbors of a point $i$, we 
$a$) translate $i \rightarrow i_{ext}$;
$b$) find $j_{ext} = i_{ext} + \Delta ij_{ext}$; and
$c$) translate $j_{ext} \rightarrow j$.
   Notice that several points of the extended mesh will map to the same point
within the unit cell and that, in principle, a unit cell point $j$ may be 
neighbor of $i$ through different values of $j_{ext}$.
   In our example, the innermost loop would then read
\begin{verbatim}
   Lf(i) = 0
   do neighb = 1,n_neighb
      j_ext = i_extended(i) + ij_delta(neighb)
      j = i_cell(j_ext)
      Lf(i) = Lf(i) + L(neighb) * f(j)
   end do
\end{verbatim}
where the number of neighbor points would be \verb'n_neighb=7', 
including the central point itself, and
\begin{verbatim}
   ij_delta(1) =  1             ; L(1) = 1/dx2
   ij_delta(2) = -1             ; L(1) = 1/dx2
   ij_delta(3) =  nx_ext        ; L(3) = 1/dy2
   ij_delta(4) = -nx_ext        ; L(4) = 1/dy2
   ij_delta(5) =  nx_ext*ny_ext ; L(5) = 1/dz2
   ij_delta(6) = -nx_ext*ny_ext ; L(6) = 1/dz2
   ij_delta(7) =  0             ; L(7) =-2/dx2-2/dy2-2/dz2 
\end{verbatim}
   Notice, however, that the above loop is completely general, for any
linear operator, using an arbitrary number of neighbor points for its
finite difference representation.
   In fact, it is even independent of the space dimensionality.
   Furthermore, the index operations required are just one 
addition and three memory calls to arrays of range one~\cite{note}.
   This inner loop simplicity comes at the expense of the two extra arrays 
\verb'i_extended' and \verb'i_cell' (of the size of the normal and extended 
meshes, respectively) which are generally an acceptable memory overhead.
   Notice, however, that the the neighbor-point list \verb'ij_delta'
is independent of the mesh index $i$, what makes this array quite small
in most problems of interest.

   Although particularly convenient for periodic boundary conditions, 
the present method may also be useful with fixed boundary conditions by,
for example, mapping the outside points of the extended mesh to a 
single memory address with a fixed boundary value.
   In conclusion, I have presented an efficient method to handle index
references in many typical problems involving periodic boundary conditions
in more than one dimension.

\acknowledgements

   This work was supported by the Fundaci\'on Ram\'on Areces and 
by MCT/DGI grant PB00-1312.


\begin{references}
\bibitem{note}
   In the present example, further savings can be achieved by extending 
the arrays \verb'f(i)' and \verb'Lf(i)' themselves, eliminating the index 
translations and facilitating the paralelization.
   Another efficient possibility is provided in Fortran90 by the
intrinsic cshift operation.
   However, these approaches are more complicated in other cases, 
like atomic-neighbor list constructions, while the double index
algorithm is quite general.
\end{references}
\end{document}